\documentclass[prb,twocolumn,floatfix,showpacs,superscriptaddress]{revtex4}
\usepackage{amsmath}
\usepackage{amssymb}
\usepackage{amsfonts}
\usepackage{graphicx}
\usepackage{dcolumn}
\usepackage{bm}
\usepackage{color}
\usepackage{chngcntr}

\DeclareGraphicsExtensions{.pdf,.png,.jpg,.eps}

\begin{document}

\title{Polar optical phonons in core-shell semiconductor nanowires}
\author{Dar\'{\i}o G. Santiago-P\'{e}rez}
\affiliation{Universidad de Sancti Spiritus ``Jos\'{e} Mart\'{\i} P\'{e}rez", Ave. de los M\'artires 360, CP 62100, Sancti Spiritus, Cuba}
\affiliation{Universidad Aut\'{o}noma del Estado de Morelos, Ave. Universidad 1001, CP 62209, Cuernavaca, Morelos, M\'{e}xico}
\author{C. Trallero-Giner}
\affiliation{Department of Theoretical Physics, Havana University, Havana 10400, Cuba}
\author{R. P\'{e}rez-\'Alvarez}
\thanks{Corresponding author}
\email{rpa@uaem.mx}
\affiliation{Universidad Aut\'{o}noma del Estado de Morelos, Ave. Universidad 1001, CP 62209, Cuernavaca, Morelos, M\'{e}xico}
\author{Leonor Chico}
\affiliation{Instituto de Ciencia de Materiales de Madrid (ICMM),
Consejo Superior de Investigaciones Cient\'{\i}ficas (CSIC),
C/ Sor Juana In\'es de la Cruz 3,
28049 Madrid, Spain}
\date{\today }

\begin{abstract}
We obtain the the long-wavelength polar optical vibrational modes of
semiconductor core-shell nanowires by means of a phenomenological continuum
model. A basis for the space of solutions is derived, and by applying the
appropriate boundary conditions, the transcendental equations for the
coupled and uncoupled modes are attained. Our results are applied to the
study of the GaAs-GaP core-shell nanowire, for which we calculate
numerically the polar optical modes, analyzing the role of strain in the
vibrational properties of this nanosystem.
\end{abstract}

\maketitle

\section{Introduction}

\label{Introduction}

The study of semiconductor nanowires is of the utmost importance for the
progress of the design and fabrication of novel devices and the
investigation of fundamental phenomena. The development of growth techniques
has allowed for the fabrication of high quality systems. Among these, the
core-shell architecture is of great interest:\cite{Peidong} a cylindrical
core of a semiconductor material is surrounded by a shell of a different
semiconductor, usually with a larger bandgap. In this way, it provides a
means of removing surface states and separating the carriers, or as a
waveguide or cavity for optoelectronic applications. Furthermore, if the
core and shell materials are grown with a lattice mismatch, the strain can
be employed as an additional degree of freedom for band structure
engineering. These particular systems have been synthesized employing
different pairs of core-shell materials, such as GaAs-GaAsP,\cite{Hua2009}
InAs-GaAs,\cite{Niquet2007} GaN-GaP,\cite{Hung-Min2003} GaP-GaN,\cite%
{Hung-Min2003} GaAs-GaP,\cite{Montazeri2010} AlN-GaN,\cite{Zhang} GaAsP-GaP,%
\cite{Mohseni}, GaAs-AlGaAs,\cite{Bailon} and CdSe/CdS,\cite{Giugni2012}
among others. A great variety of applications for these core-shell nanowires
have appeared, for instance, nanowire lasers,\cite{Hua2009} nanowire
nanosensors,\cite{Cui,Zheng} photovoltaic devices\cite{Dong} and light
emission diodes,\cite{Minot} to name a few.

Polar optical phonons are of a great interest for the spectroscopic
characterization of core-shell nanowires of compound semiconductors. Raman
scattering provides information on the phonon frequencies, which can be
related to the strain in the core and the shell of the nanowires. However,
in spite of its importance for their spectroscopic characterization,
up to our knowledge, only a few calculations of interface modes
in GaN/AlN core-shell nanowires have been 
reported recently, mainly obtained by means of a 
macroscopic dielectric model.~\cite{cross}

In this work we address this issue, employing a phenomenological continuum
model for polar optical phonons in the long-wave limit in a cylindrical
core-shell geometry. Indeed, polar optical oscillations have been
successfully studied for different nanostructures applying a long-wavelength
approximation and based on different continuum approaches; see, for example,
Refs.~\onlinecite{CubaLibro, StroscioDutta2001, ridley} and references
therein. In particular, oscillations in cylindrical systems have been
studied in Refs.~\onlinecite{Stroscio, Enderlein, Comas1995}, but only for
solid nanowires made of a single material, and in some cases neglecting the
dispersion along the nanowire axis.

In order to study polar phonons in core-shell nanowires, we follow the
approach employed for other geometries, as outlined in Refs.~\onlinecite{CubaLibro, Comas1995},
and originally exposed in Ref.~\onlinecite{CTG92}. This phenomenological continuum model (PCM) takes into
account the coupled electro-mechanical character of the vibrations without
making any simplifying assumptions. From the prior experience
in quasi-two-dimensional \cite{CubaLibro} and quasi-zero-dimensional
systems,\cite{Roca1994} as it was shown in Refs.~\onlinecite{CTG92,CTG94,Comas94},
we do know that no further hypotheses as
to the electromechanical coupling should be made when searching for linearly
independent solutions in these quasi-one-dimensional structures. Some
previous work following this approach has been made for cylindrical
geometries,\cite{Comas1995} albeit for simple (i.e., solid and with only one
material) nanowires and without considering the dependence on the axial
coordinate. Here we generalize this work, considering 
all types of polar oscillation modes in core-shell nanowires. To this end we
focus in obtaining a basis function with cylindrical symmetry, taking into
account the possible angular and axial dependence the modes may have. We
apply the appropriate boundary conditions for core-shell modes to a general
solution, given by a linear combination of the basis functions. As we
concentrate in materials with very different bulk values of their
mechanical parameters, we can impose a total confinement of the mechanical
components. This condition leads to the mixing of the different modes.
Indeed, phonon modes of mixed nature are obtained and may display
predominant longitudinal optical ($LO$), transverse optical ($TO$) or interface
($I$) profiles in the different regions of the vibrational spectra. We analyze
the character of the phonon modes, and give detailed numerical results for
one particular case, namely, the GaAs-GaP core-shell nanowire.

The paper is organized as follows: In Sec. II we present the fundamental
equations which describe the polar oscillation modes, discussing their
physical meaning and obtaining a basis for the cylindrical geometry. Sec.
III explains the obtention of the polar optical modes in core-shell
nanowires by applying the appropriate boundary conditions. The solution for
the interface optical modes for the core-shell nanowires with cylindrical
cross section in the framework of the dielectric continuum model is
presented in Sec. IV. Sec. V discusses the inclusion of strain effects in
our model. In Sec. VI the results corresponding to a particular example, the
GaAs-GaP core-shell nanowire, are presented. In Sec. VII we draw our
conclusions.

\section{\label{Modelo}The phenomenological continuum model in cylindrical
coordinates}

We briefly recall here the formalism of the PCM employed in this work.
Following the procedure developed in Refs.~\onlinecite{CTG92,CTG94,Comas94},
the fundamental equations of motion which include the bulk phonon dispersion are given by

\begin{equation}
\rho _{m}(\omega ^{2}-\omega _{TO}^{2})\vec{u}=\rho _{m}\beta _{L}^{2}\nabla
(\nabla \cdot \vec{u})-\rho _{m}\beta _{T}^{2}\nabla \times \nabla \times
\vec{u}+\alpha \nabla \varphi ,  \label{ecuacion de movimiento2}
\end{equation}%
and

\begin{equation}
\nabla ^{2}\varphi =\frac{4\pi \alpha }{\varepsilon _{\infty }}\nabla \cdot
\vec{u},  \label{potencial}
\end{equation}

\noindent with the parameter $\alpha$ defined as

\begin{eqnarray}  \label{alfa}
\alpha^{2}=\frac{(\varepsilon_{0}-\varepsilon_{\infty})\rho_{m}%
\omega^{2}_{TO}}{4\pi}.
\end{eqnarray}

In these expressions, $\omega_{TO}$ is the transversal bulk frequency at the
$\Gamma$ point, $\rho_{m}$ is the reduced mass density, $\beta_{L}$ ($%
\beta_{T}$) describes the quadratic dispersion of the $LO$ ($TO$)-bulk phonon
dispersion of the optical modes in the long-wave limit, and $\varepsilon_{0}$
($\varepsilon_{\infty}$) is the static (high frequency) dielectric constant.
The relative mechanical displacement of the ions is represented by $\vec{u}$
and the electric potential due to the polar character of the vibrations is
denoted by $\varphi$. In this model the equations are treated in the
quasi-stationary approximation so a harmonic time dependence is considered
for all the involved quantities.

\begin{figure}[htb]
\includegraphics[width=\columnwidth]{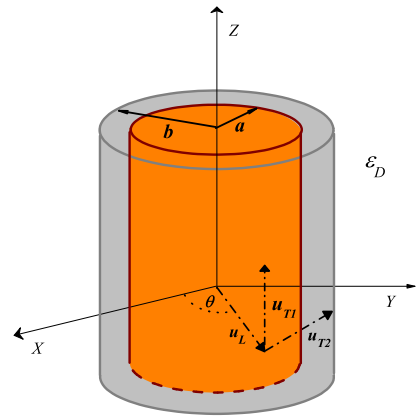}
\caption{(Color online) Schematic representation of the capped wire system
under study. In our case, the material in the core is GaAs and GaP in the
shell. $a$ ($b$) is the core (shell) radius. Vibrational phonon amplitudes
$u_{L}$, $u_{T1}$, and $u_{T2}$ are indicated. The effect associated to the
embedding matrix on the vibrational modes is characterized by an outer
dielectric constant $\protect\varepsilon_{D}$.}
\label{Fig1}
\end{figure}

Eqs.~(\ref{ecuacion de movimiento2}) and (\ref{potencial}) represent a
system of four coupled partial differential equations which describe the
confined polar optical phonons in each region of the semiconductor
heterostructure. In this particular case, hybrid core-shell cylindrical
nanowires consist of a material ``$s$" grown on a core structure of material
``$c$" , and the medium properties are considered piecewise, as depicted in
Fig.~\ref{Fig1}. Furthermore, the nanowire is embedded in a host material,
which is typically a silicate matrix or an organic polymeric compound.

We model the core-shell nanowire as an infinite cylinder of circular cross
section with radius $a$, dressed by a cylindrical shell of another material
with external radius $b$ (see figure \ref{Fig1}). The wire is embedded in a
host material uncoupled to the oscillations of the nanowire, characterized
by its dielectric constant $\varepsilon_{D}$.

In order to find a general solution for the oscillations of the nanowire, we
have to find a basis for the solutions in each region.
With this purpose, we follow the method of the potentials described in
detail in the book by Morse and Feshbach,\cite{Morse} that we sketch briefly.

First, we introduce the auxiliary potentials $\vec{\Gamma}$ and $\Lambda$
such that

\begin{eqnarray}  \label{auxiliares}
\vec{\Gamma}=\nabla\times\vec{u}\;\;\; \text{and} \;\;\;\Lambda=\nabla\cdot%
\vec{u}\;.
\end{eqnarray}

Taking the curl and the divergence of Eq. (\ref{ecuacion de movimiento2}),
we obtain the following new equations for the potentials:

\begin{eqnarray}
\nabla ^{2}\vec{\Gamma}+Q_{T}^{2}\vec{\Gamma} &=&\vec{0},  \label{Gamma} \\
\nabla ^{2}\Lambda +Q_{L}^{2}\Lambda &=&0,  \label{Lambda1}
\end{eqnarray}

\noindent
with $Q_{T},Q_{L}$ given by

\begin{eqnarray}  \label{nuevas ecuaciones}
Q^{2}_{T}=\frac{\omega^{2}_{TO}-\omega^{2}}{\beta^{2}_{T}}, \\
Q^{2}_{L}=\frac{\omega^{2}_{LO}-\omega^{2}}{\beta^{2}_{L}}.  \notag
\end{eqnarray}

It can be seen that the solution of
Eq.~(\ref{potencial}) can be written as
\begin{equation}
\varphi=\varphi_{H}-\frac{4\pi\alpha}{\varepsilon_{\infty}Q^{2}_{L}}\Lambda,
\notag
\end{equation}
\noindent where $\varphi_{H}$ is the solution of the Laplace equation $%
\nabla^{2}\varphi_{H}=0$. Moreover, straightforward mathematical
manipulations lead us to the following expression for $\vec{u}$:

\begin{eqnarray}  \label{vector desplazamiento}
\vec{u}=-\nabla\left[\frac{\alpha}{\rho_{m}\beta^{2}_{T}Q^{2}_{T}}%
\varphi_{H}+ \frac{\Lambda}{Q^{2}_{L}}\right]+ \frac{1}{Q^{2}_{T}}%
\nabla\times\vec{\Gamma}.
\end{eqnarray}

In order to obtain the general solution for the relative mechanical
displacement $\vec{u}$ and the electric potential $\varphi_{H}$ it is
necessary to solve the Helmholtz equations for $\Lambda$ and $\vec{\Gamma}$
and the Laplace equation for $\varphi_{H}$. The solutions for $\Lambda$ and $%
\varphi_{H}$ are obtained in the standard way; the method for
solving the Helmholtz vectorial equation (\ref{Gamma})\ in cylindrical
coordinates is reported\ in Refs.~\onlinecite{Ruppin,Morse}. Nevertheless,
in the general case of cylindrical symmetry the displacement vector (\ref%
{vector desplazamiento}) cannot be decoupled into two independent directions
with 
pure longitudinal ($L$) or transversal motion ($T$). As we state below, only
under particular conditions we are able to decouple the motion in $L$ and $T$
independent oscillations. 
It is convenient to express the vector potential $\vec{\Gamma}$ as a linear
combination of the vectors $\vec{M}$ and $\vec{N}$,

\begin{eqnarray}  \label{auxiliares Gamma}
\vec{M}&=&\nabla\times(\upsilon_{1}\vec{\kappa}), \\
\vec{N}&=&\frac{1}{Q^{2}_{T}}\nabla\times\nabla\times(\upsilon_{2}\vec{\kappa%
}),  \notag
\end{eqnarray}

\noindent where $\upsilon_{i}, (i=1,2)$ are linearly independent solutions
of the scalar equation $\nabla^{2}\upsilon_{i}+Q^{2}_{T}\upsilon_{i}=0$ and,
in cylindrical geometry, $\vec{\kappa}=\vec{e}_{z}$.

Once the functions $\Lambda$, $\vec{\Gamma}$ and $\varphi_{H}$ have been
obtained, it is easy to prove that the general solution of Eqs.~(\ref{ecuacion de movimiento2})
can be expressed in terms of the analytical basis
functions for the space of solutions, given by

\begin{eqnarray}  \label{vectoresbase}
F_{T1}&=&\left(
\begin{array}{c}
\vec{u}_{T1} \\
\varphi_{T1} \\
\end{array}
\right)=\left(
\begin{array}{c}
\frac{ik_{z}}{q_{T}}f_{n}^{\prime }(q_{T}\rho) \\
-\frac{nk_{z}}{q_{T}}\frac{1}{q_{T}\rho}f_{n}(q_{T}\rho) \\
f_{n}(q_{T}\rho) \\
0 \\
\end{array}
\right)e^{i(n\theta+k_{z}z)}  \notag \\
F_{T2}&=&\left(
\begin{array}{c}
\vec{u}_{T2} \\
\varphi_{T2} \\
\end{array}
\right)=\left(
\begin{array}{c}
\frac{in}{q_{T}\rho}f_{n}(q_{T}\rho) \\
-f^{\prime }_{n}(q_{T}\rho) \\
0 \\
0 \\
\end{array}
\right)e^{i(n\theta+k_{z}z)}  \notag \\
F_{L}&=&\left(
\begin{array}{c}
\vec{u}_{L} \\
\varphi_{L} \\
\end{array}
\right)=\left(
\begin{array}{c}
f^{\prime }_{n}(q_{L}\rho) \\
\frac{in}{q_{L}\rho}f_{n}(q_{L}\rho) \\
\frac{ik_{z}}{q_{L}}f_{n}(q_{L}\rho) \\
\frac{4\pi\alpha}{\varepsilon_{\infty}}\frac{1}{q_{L}}f_{n}(q_{L}\rho)%
\end{array}
\right)e^{i(n\theta+k_{z}z)}  \label{basic} \\
F_{H}&=&\left(
\begin{array}{c}
\vec{u}_{H} \\
\varphi_{H} \\
\end{array}
\right)=\left(
\begin{array}{c}
g^{\prime }_{n}(k_{z}\rho) \\
\frac{in}{k_{z}\rho}g_{n}(k_{z}\rho) \\
g_{n}(k_{z}\rho) \\
-\frac{\rho_{m}\beta^{2}_{T}q^{2}_{T}}{\alpha}\frac{1}{k_{z}}g_{n}(k_{z}\rho)%
\end{array}
\right)e^{i(n\theta+k_{z}z)},  \notag
\end{eqnarray}

\noindent where the matrix components are understood in the form $(u_{\rho},
u_{\theta}, u_{z}, \varphi)$; the prime denotes the derivative with respect
to the argument; $n$ is an integer label related to the angular dependence
of the modes; and  $k_z$ is the continuum
wavevector along the cylinder axis. Additionally, we introduce the
wavenumbers

\begin{eqnarray}
q^{2}_{L,T} &=& Q^{2}_{L,T} -k_z^{2} \;.
\end{eqnarray}

If $q^{2}_{L,T}>0$ ($q^{2}_{L,T}<0$) the function $f_{n}$ is an order-$n$
Bessel (modified Bessel) function of the first or second kind, i.e., Bessel $%
J_{n}$ or Neumann $N_{n}$ (Infeld $I_{n}$ or MacDonald $K_{n}$). On the
other hand, $g_{n}$ is an order-$n$ modified Bessel function of the first or
second kind, i.e., Infeld $I_{n}$ or MacDonald $K_{n}$. We follow the
definitions and conventions of Abramowitz and Stegun.~\cite{AbramowitzStegun1964}
It is important and straightforward to check that $%
\nabla\times\vec{u}_{L} = \nabla\times\vec{u}_{H} = \vec{0}$ and $\nabla\cdot%
\vec{u}_{T1} = \nabla\cdot\vec{u}_{T2} = \nabla\cdot\vec{u}_{H} = 0$.

Particular cases of this basis have been used to study phonon modes in
non-polar nanotubes \cite{ChicoRPA2004,CPC2006} and in solid nanowires with
only one material at $k_{z}=0$.~\cite{Comas1993a}

\section{Polar optical oscillation modes in core-shell nanowires}

\label{core-shell}

In order to obtain the particular solution for the core-shell geometry,
boundary conditions for $\vec{u}$ and $\varphi$ at each interface should be
applied. With respect to the electromagnetic magnitudes, we have that the
electric potential $\varphi$ and the normal component of the displacement
field $\vec{D}$ should be continuous at the interfaces. Recall that the
electric displacement vector is given by\cite{CubaLibro} $\vec{D} =
4\pi\alpha\vec{u}-\varepsilon_{\infty}\nabla\cdot\varphi$. However, we will
consider pairs of core-shell materials with disparate mechanical properties,
so the oscillations occurring in one of them do not penetrate significantly
into the other. This is the case of the GaAs-GaP core-shell nanowire
discussed in this work; in fact, a significant number of pairs of materials
of current interest satisfies this requisite. With this assumption, it can
be adopted an approximate boundary condition of complete mechanical
confinement, $\vec{u}|_{S} = 0$. Thus, the matching boundary conditions are
reduced to

\begin{eqnarray}  \label{matchingcond}
\left. \vec{u} \right|_{S}&=&0,  \notag \\
\left. \varphi^{-}\right|_{S}&=&\left. \varphi^{+}\right|_{S}, \\
\left. \varepsilon^{-}_{\infty}\frac{\partial\varphi^{-}}{\partial\rho}
\right|_{S}&=&\left. \varepsilon^{+}_{\infty}\frac{\partial\varphi^{+}}{%
\partial\rho}\right|_{S}.  \notag
\end{eqnarray}

In Eqs.~(\ref{matchingcond}) the symbol $-$(+) represents that the
associated quantity is evaluated at the inside (outside) of the
corresponding interfaces, namely, the cylindrical surfaces of radius $a$ and
$b$. The dispersion relations are then obtained applying these boundary
conditions (\ref{matchingcond}) to a general linear combination of the basis
functions, that can be written as

\begin{eqnarray}  \label{lincomb}
F=\left\{
\begin{array}{cc}
\sum\limits_{M} A_M^{(c)} F_M^{(1)} ; & \rho \leq a \\
\sum\limits_{M} A_M^{(s)} F_M^{(1)} + \sum\limits_{M}B_M^{(s)} F_M^{(2)}; &
a \leq \rho \leq b \\
\sum\limits_{M} B_M^{(D)} F_M^{(2)}; & \rho \geq b%
\end{array}%
\right.  \label{genesol}
\end{eqnarray}

\noindent where $M=T1,T2, L, H$ and $i=1,2$ denotes that the corresponding
Bessel and modified Bessel functions $f_n$, $g_n$ appearing the basis
functions $F_M^{(i)}$ are or the first or the second kind respectively.

\section{Interface optical phonons}

The system of Eqs.~(\ref{ecuacion de movimiento2}) and (\ref{potencial})
lead to coupled modes at the interfaces. These modes shows a predominant
electric character associated to the system interfaces and are related to
interface phonons (IP). For a simple characterization and for sake of
comparison with the present theoretical model, we calculate the IP employing
the dielectric continuum approach (DCA). Considering that the electric field
satisfies quasistatic Maxwell equations, we have
\begin{equation*}
\varepsilon _{c(s)}(\omega )\nabla ^{2}\varphi =0,
\end{equation*}%
where the frequency dependent dielectric function $\varepsilon
_{c(s)}(\omega )$ for the core (shell) is given by the standard expression
\begin{equation}
\varepsilon _{c(s)}(\omega )=\varepsilon _{\infty }^{c(s)}\frac{\omega
_{LO}^{c(s)2}-\omega ^{2}}{\omega _{TO}^{c(s)2}-\omega ^{2}}.
\end{equation}%
In the above equation $\omega _{LO}^{c(s)}$ and $\omega _{TO}^{c(s)}$ are
the bulk longitudinal and transversal polar optical phonons frequencies at
the $\Gamma $ point for the core (shell) semiconductor material. The IP
satisfy the Laplace equation $\nabla ^{2}\varphi =0$ and $\varepsilon
_{c(s)}(\omega )\neq 0.$ Thus, employing the standard electrostatic boundary
condition at the interfaces we obtain
\begin{widetext}
\begin{eqnarray}
(\varepsilon _{c}(\omega )-\varepsilon _{s}(\omega ))(\varepsilon
_{D}-\varepsilon _{s}(\omega ))I_{n}(k_{z}a)I_{n}^{\prime
}(k_{z}a)K_{n}(\gamma k_{z}a)K_{n}^{\prime }(\gamma k_{z}a) &-&  \notag \\
(\varepsilon _{c}(\omega )K_{n}(k_{z}a)I_{n}^{\prime }(k_{z}a)-\varepsilon
_{s}(\omega )I_{n}(k_{z}a)K_{n}^{\prime }(k_{z}a)) &\times &  \notag \\
(\varepsilon _{d}I_{n}(\gamma k_{z}a)K_{n}^{\prime }(\gamma
k_{z}a)-\varepsilon _{s}(\omega )K_{n}(\gamma k_{z}a)I_{n}^{\prime }(\gamma
k_{z}a)) &=&0,  \label{IF}
\end{eqnarray}
\end{widetext}where $\gamma =b/a$ is the ratio between the shell and core
radii. Equation~(\ref{IF}) gives the dispersion relations of IP as a
function of 
$k_{z}$ and the parameter $\gamma $ for different values of $n=0,1,2,..$. In
the case of $k_{z}=0$ Eq.~(\ref{IF}) is reduced to
\begin{widetext}
\begin{equation*}
(\varepsilon _{c}(\omega )-\varepsilon _{s}(\omega ))(\varepsilon
_{D}-\varepsilon _{s}(\omega ))-(\varepsilon _{c}(\omega )+\varepsilon
_{s}(\omega ))(\varepsilon _{D}+\varepsilon _{s}(\omega ))\gamma ^{2n}=0.
\end{equation*}
\end{widetext}
Equation~(\ref{IF}) shows that for each value of $n$ we have three
independent IP branches. One is linked to the cylindrical core embedded in a
host material with an effective dielectric constant, and the other two
correspond to the cylindrical shell structure sandwiched between the core
and a host dielectric medium. These interface phonons depend on the
geometrical parameter $\gamma $. 

\section{Strain effects}

It is important to note that in semiconductor core-shell nanowires, strain
effects cannot be neglected. To model these, we applied the same procedure
as in Ref.~\onlinecite{paperanterior}, and we study its importance by
comparing to the strain-free case.

The shift in the optical phonon frequencies due to the strain in the
core-shell nanowire 
is given by\cite{Spanier,Dohcevic}

\begin{eqnarray}  \label{corrimiento}
\Delta\omega_{i}=-\gamma_{i}\omega_{i}\frac{\Delta V}{V},
\end{eqnarray}

\noindent where $i = LO,TO$, $\gamma_{i}$ is the Gr\"uneisen parameter,
$V$ is the volume of unit cell, and $\Delta V$ is the volume change due to
the lattice mismatch. The relation $\frac{\Delta V}{V}=\mathrm{tr}
(\varepsilon)$, where $\mathrm{tr}(\varepsilon)$ is the trace of the stress
tensor, can be evaluated for the core and shell materials in cylindrical
geometry,\cite{Menendez} yielding

\begin{widetext}
\begin{eqnarray}
{\rm tr}(\varepsilon_{c})&=&-2\varepsilon_{\rm misfit}\left(\frac{(1+\nu_{c})(1-2\nu_{c})(\gamma^{2}-1)}{(1-E_r)(1-2\nu_{c})-(1-2\nu_{c}+E_r)\gamma^{2}}+\frac{\nu_{c}(\gamma^{2}-1)}{(\gamma^{2}-1)
+E_r}\right)+\varepsilon_{\rm misfit}\left(\frac{(\gamma^{2}-1)}{(\gamma^{2}-1)+E_r}\right),
\label{core} \\
{\rm tr}(\varepsilon_{s})&=&2\varepsilon_{\rm misfit}\left(\frac{(1+\nu_{s})(1-2\nu_{s})E_r}{(1-E_r)(1-2\nu_{s})-(1-2\nu_{s}+E_r)\gamma^{2}}+\frac{\nu_{s}E_r}{(\gamma^{2}-1)+E_r}\right)
-\varepsilon_{\rm misfit}\frac{E_r}{(\gamma^{2}-1)+E_r},
\label{shell}
\end{eqnarray}
\end{widetext}

\noindent with $\nu_{c}$, $\nu_{s}$ the Poisson ratios of the core and shell
materials respectively; the lattice mismatch is $\varepsilon_{\mathrm{misfit}%
}=(a_{s}-a_{c})/a_{c}$ with $a_{c}$ $(a_{s})$ being the cubic lattice
constant of the core (shell) material; $E_r=E_c/E_s$ is the ratio between
the core and shell Young moduli.
From Eqs.~(\ref{core}) and (\ref{shell}) we obtain the following limits
\begin{eqnarray}  \label{limtrauno}
\mathrm{tr}(\varepsilon _{c})_{\lim \gamma \rightarrow 1}&=&0; \\
\ \ \mathrm{tr}(\varepsilon _{s})_{\lim \gamma \rightarrow 1}&=&2\varepsilon
_{\mathrm{misfit}}\frac{1-2\nu _{c}}{\nu _{c}-1};  \notag \\
\mathrm{tr}(\varepsilon _{c})_{\lim \gamma \rightarrow \infty
}&=&\varepsilon_{\mathrm{misfit}}\frac{(1-2\nu _{c})\left( 3+\nu _{c}\right)
}{1-2\nu _{c}+E_{r}}; \\
\mathrm{tr}(\varepsilon _{s})_{\lim \gamma \rightarrow \infty }&=&0.  \notag
\label{limtrain}
\end{eqnarray}
These equations allow us to include strain effects on the phonon frequencies
in our model, replacing in Eqs.~(\ref{nuevas ecuaciones}) and (\ref{IF}) the
unstrained bulk frequencies at $\Gamma$, $\omega_{TO}$ and $\omega_{LO}$, by
$\omega_{T}(\gamma)$, $\omega_{L}(\gamma)$:

\begin{eqnarray}  \label{efecto del esfuerzo}
\omega_{T}(\gamma)&=&\omega_{TO}+\Delta\omega_{TO}(\gamma); \\
\omega_{L}(\gamma)&=&\omega_{LO}+\Delta\omega_{LO}(\gamma).
\end{eqnarray}

\section{Results}

In what follows we present some analytical and numerical results for core
and shell modes, without and with stress effects. As commented above, we
choose as a representative example the GaAs/GaP core-shell nanowire. The
parameters chosen for these two materials are listed in Table \ref%
{Parametros}. Since the $k_z=0$, $n=0$ case has been analyzed elsewhere,
\cite{paperanterior} we focus on $k_z = 0$, $n>0$ and $k_z \neq 0$, $n=0,1$
cases. In order to avoid a heavy notation, we drop the indices $c$, $s$ in
the parabolicity parameters and in the bulk frequencies when there is no
possible ambiguity.\newline

\begin{widetext}
\begin{center}
\begin{table}[htb]
\caption{\label{Parametros} Bulk parameters for GaAs and GaP in zinc blende phase.}
\begin{tabular}{cccccccccccc}
  \hline
  \hline
  Material & $\epsilon_{0}$ & $\epsilon_{\infty}^{*}$ & $\omega_{TO}$ (cm$^{-1}$) & $\omega_{LO}$ (cm$^{-1}$) & $\beta_{T} (\times10^{-6})$ & $\beta_{L} (\times10^{-6})$ & $\gamma_{TO}$ & $\gamma_{LO}$ & $E$ (10$^{12}$ dyn/cm$^{2}$) & $\nu$ & $a_{0}$ (nm)       \\
  \hline
  \hline
  GaAs     & 12.80$^{a}$    & 11.26      & 267$^{a}$                 & 285$^{a}$                 & 1.70$^{b}$                  & 1.76$^{b}$ & 1.11$^{d}$ & 0.97$^{d}$ & 0.853$^{d}$ & 0.312$^{d}$ & 0.565$^{d}$ \\
  GaP      & 11.11$^{a}$    &  9.15     & 365.3$^{a}$               & 402.5$^{a}$               & 0.72$^{c}$                  & 1.60$^{c}$ & 1.09$^{d}$ & 0.95$^{d}$ & 1.03$^{d}$  & 0.306$^{d}$ & 0.545$^{d}$ \\
  \hline
  \hline
\end{tabular}

a) Ref. \onlinecite{Martienssen}; b) Ref. \onlinecite{CubaLibro}; c) Ref. \onlinecite{Borcherds}; d) Ref. \onlinecite{Adachi}.\\
$^*$Using the Lyddane-Sachs-Teller relation.
\end{table}
\end{center}
\end{widetext}

\subsection{\label{subsec:nneq0k0}Modes with $n\neq0$, $k_z=0$}

\subsubsection{\label{SubsectionCoreModes}Core modes}

Assuming complete mechanical confinement, we model core modes by considering
$\vec{u}\equiv0$ for $a<\rho<b$ and $\vec{u}\neq0$ for $\rho<a$. The
application of the boundary conditions indicated in Eqs.~(\ref{matchingcond})
yields one family of uncoupled $T1$ modes and one of coupled $L$-$T2$
modes. This decoupling of the $T1$ modes is evident from the expressions of
the basis functions (\ref{vectoresbase}) for $k_z=0$. The eigenvalue
equations for the uncoupled $T1$ modes are given by $J_{n}(\mu^{(m)}_{n})=0,
m=1,2,...$ which yield the dispersion relations $\omega^{2}=%
\omega_{TO}^{2}-(\mu^{(m)}_{n}\beta_{T}/a)^{2}$.


\begin{figure}[htb]
\includegraphics[width=\columnwidth]{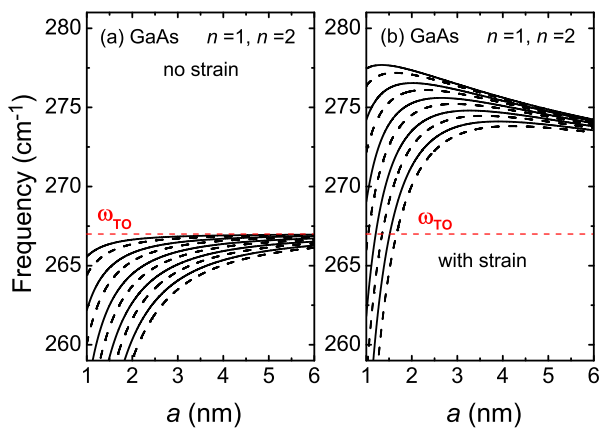} 
\caption{(Color online) GaAs optical uncoupled transversal phonon modes at $%
k_{z}=0$ for $n=1$ (full lines) and $n=2$ (dashed lines) in a GaAs-GaP
core-shell nanowire as a function of the core radius $a$. Panel (a) without
strain; panel (b)including strain effects. For the calculation represented
in panel (b) we fixed $b-a=3$ nm. The bulk GaAs TO phonon frequency is
indicated by a red (gray) horizontal dashed line.}
\label{modosdesacopladoscoren12}
\end{figure}

\begin{figure}[htb]
\includegraphics[width=\columnwidth]{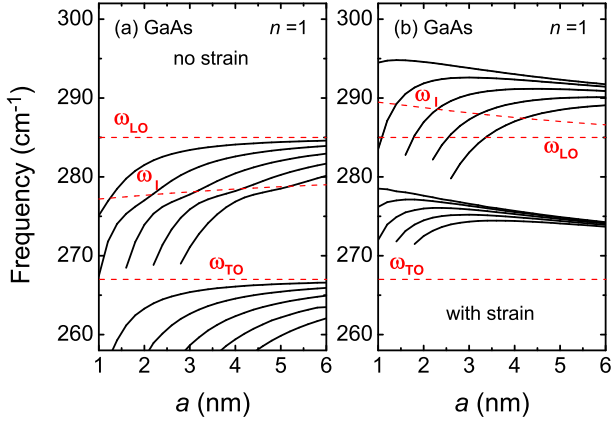} 
\caption{(Color online) GaAs optical coupled phonon modes at $k_{z}=0$ for $%
n=1$ in GaAs-GaP core-shell nanowire as a function of the core radius $a$.
Panel (a) neglecting strain effects, panel (b) including strain. In the
calculation we fixed the value of $b-a=3$ nm and $\protect\varepsilon%
_{D}=2.56$. The bulk GaAs $LO$ and $TO$ phonon frequencies are indicated by
gray (red) dashed lines. The corresponding interface phonon frequency
obtained in the framework of the DCA is also represented by a gray (red)
dashed line.}
\label{modosacopladoscoreespesorfijon1}
\end{figure}

Figure \ref{modosdesacopladoscoren12} shows the frequency dependence
 of the confined modes on the core radius $a$ for the GaAs-GaP
core-shell nanowire. 
For the strain-free case (Fig.~\ref{modosdesacopladoscoren12} (a)), the mode
frequency is independent of the shell radius $b$, so it is similar to an
undressed quantum wire. This behavior changes when the effects of strain are
taken into account, which yields the eigenfrequencies dependent on the shell
radius $b$. As in the $n=0$ case, there is an increase on the frequencies of
the modes when considering strain effects, clearly shown in Fig.~\ref{modosdesacopladoscoren12} (b).
Following the results of the Appendix for
the coupled $L$-$T2$ core modes, the secular equation~(\ref{modosLT2shellk0}) is reduced to the following:

\begin{widetext}
\begin{eqnarray}
\label{modosLT2corek0}
\left[J'_{n}(t_{c})-\frac{n}{t_{c}}J_{n}(t_{c})\right]\left[J'_{n}(l_{c})-C_{1}\frac{n}{l_{c}}J_{n}(l_{c})\right]-C_{2}\left[1-\frac{\omega^{2}}{\omega_{TO}^{2}}\right]\left[J'_{n}(t_{c})J'_{n}(l_{c})-\frac{n^{2}}{t_{c}l_{c}}J_{n}(t_{c})J_{n}(l_{c})\right]=0,
\end{eqnarray}

\noindent where

\begin{eqnarray}
C_{1}&=&\frac{\varepsilon^{s}_{\infty}[(\varepsilon_{D}-\varepsilon^{s}_{\infty})+\gamma^{2n}(\varepsilon_{D}+\varepsilon^{s}_{\infty})]}{\varepsilon^{c}_{\infty}[(\varepsilon_{D}-\varepsilon^{s}_{\infty})-\gamma^{2n}(\varepsilon_{D}+\varepsilon^{s}_{\infty})]},
\\
C_{2}&=&\frac{[(\varepsilon^{s}_{\infty}-\varepsilon^{c}_{\infty})(\varepsilon_{D}-\varepsilon^{s}_{\infty})+\gamma^{2n}(\varepsilon^{c}_{\infty}+\varepsilon^{s}_{\infty})(\varepsilon_{D}+\varepsilon^{s}_{\infty})]}{(\varepsilon^{c}_{0}-\varepsilon^{c}_{\infty})[(\varepsilon_{D}-\varepsilon^{s}_{\infty})-\gamma^{2n}(\varepsilon_{D}+\varepsilon^{s}_{\infty})]},
\;\nonumber
\end{eqnarray}
\end{widetext}and $t_{c}=q_{T}a$, $l_{c}=q_{L}a$. Sub- or superindices $c$, $%
s$, indicate that the corresponding quantities (i.e., $q_{T,L}$ or the
dielectric constants $\varepsilon _{0}$, $\varepsilon _{\infty }$)
correspond to the core or shell materials respectively.

The phonon frequencies for $n=1$ as a function of core radius $a$, given by
Eq.~(\ref{modosLT2corek0}), are presented in Fig.~\ref{modosacopladoscoreespesorfijon1}.
The interface ($I$) mode manifests in the
abrupt change of slope in the frequencies, where the mixing between
longitudinal and transversal modes occurs. The electrostatic potential of
the surface oscillation is manifested when the interaction of the $LO$%
-confined phonon with the surface mode becomes strong for certain values of
the core radius $a$. In this region the electric character of the modes is
dominant.
As $a\rightarrow\infty$, the bulk $LO$ and $TO$ phonon dispersion relations
are recovered. The effect of strain is also an increase of the phonon
frequencies (Fig. \ref{modosacopladoscoreespesorfijon1} (b)).
Notice the characteristic change of the slope
in Figs.~\ref{modosdesacopladoscoren12} (b) and \ref{modosacopladoscoreespesorfijon1} (b)
when the strain is considered. As the
core radius increases, the parameter $\gamma\rightarrow 1$ and the phonon
frequencies decrease, reaching the bulk limit $\omega_{T(L)}(\gamma=1)=%
\omega_{TO(LO)}$; then, the spatial confinement and the influence of the
strain (see Eq.~(\ref{limtrauno})) on the core are negligible.

\begin{figure}[htb]
\includegraphics[width=\columnwidth]{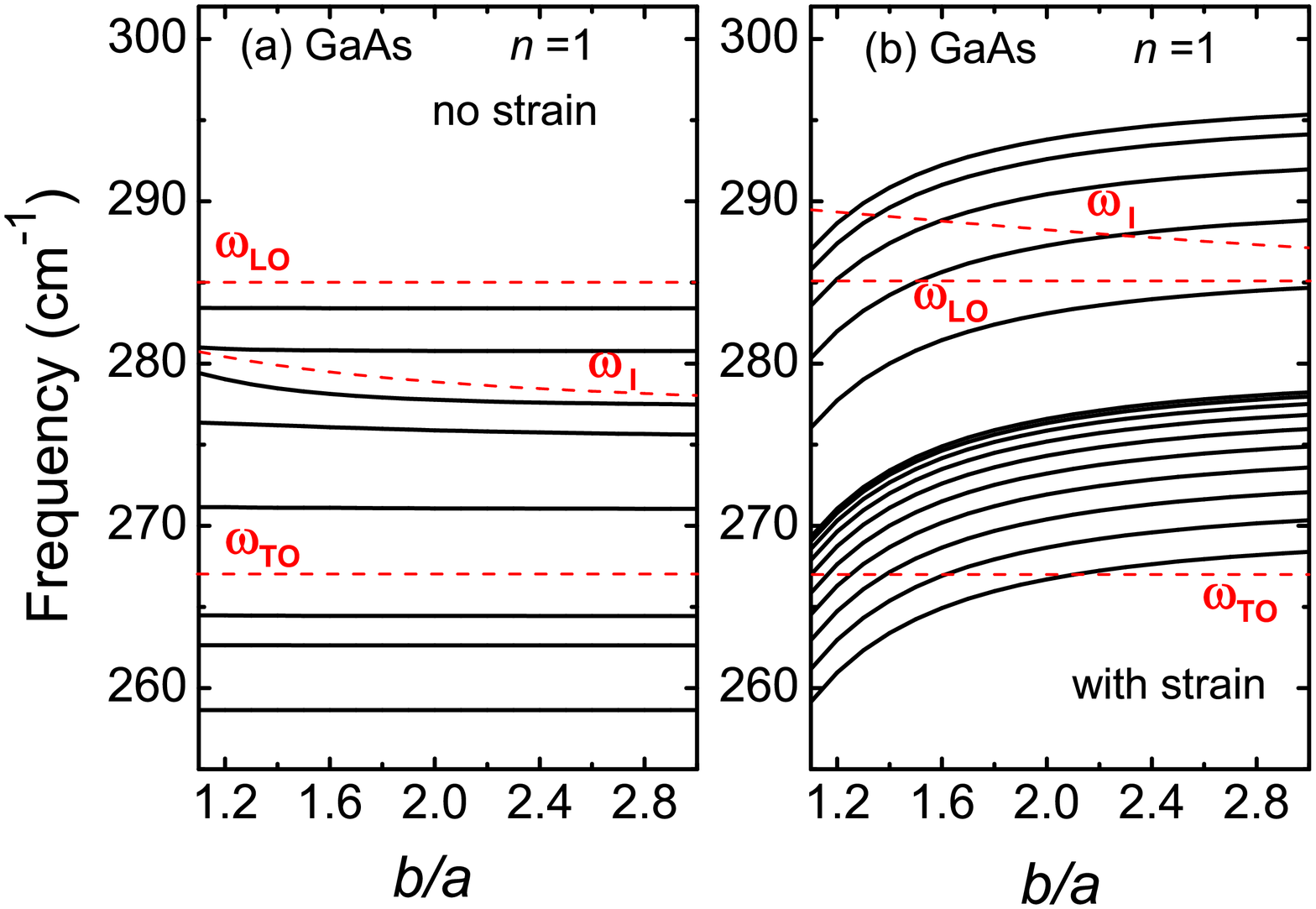} 
\caption{(Color online)
The same asFig.~(\ref{modosacopladoscoreespesorfijon1}) as a function of the ratio $b/a$,
fixing the value of $a=3$ nm and$\protect\varepsilon_{D}=2.56$.
%
The corresponding interface optical phonon ($I$-phonon) frequency
obtained from Eq.~(\protect\ref{IF}) is
also represented by a gray (red) dashed line.}
\label{modosacopladoscoreradiofijon1}
\end{figure}

Figure \ref{modosacopladoscoreradiofijon1} shows the coupled core modes at $%
k_{z}=0$ for $n=1$ as a function of the ratio $b/a$. As it can be seen, the
core modes depend very weakly on the ratio $b/a$ if strain effects are
neglected (Fig. \ref{modosacopladoscoreradiofijon1} (a)). Only the third
mode in order of decreasing frequency shows a certain dependence on $\gamma$.
This is due to the mixture between longitudinal and transversal modes near
the interface phonon frequency $\omega_I$ produced by the strong interaction
of the electrostatic potential with the interface phonon (see Fig.~\ref%
{modosacopladoscoreradiofijon1}(a)). When strain effects are taken into
account the $\gamma$ dependence is governed mainly by them, producing an
upward shift of the mode frequencies Fig. \ref{modosacopladoscoreradiofijon1}%
(b). When the shell thickness is much larger than the core radius, $%
\gamma\gg1$, the phonon modes feel a residual stress described
by Eq.~(\ref{limtrain}), so the frequencies of the transversal ($\omega
_{T}(\gamma \rightarrow \infty)$) and longitudinal modes ($\omega
_{L}(\gamma \rightarrow \infty)$) reach the limit value 279.54 cm$^{-1}$ and
296.70 cm$^{-1}$ respectively.

\subsubsection{\label{SubsectionShellModes}Shell modes}

Complete mechanical confinement in the shell amounts to $\vec{u}\equiv 0$
for $\rho<a$ but $\vec{u}\neq 0$ for $a<\rho<b$. As in the core case, the
basis functions (\ref{vectoresbase}) when $k_z=0$ show that the mode $T1$ is
decoupled from the rest, while there is $L$-$T2$ coupling. Indeed, the
matching boundary conditions (\ref{matchingcond}) lead to one family of
uncoupled $T1$ modes and one of coupled $L$-$T2$ modes, that we describe in
the following.

\begin{figure}[htb]
\caption{(Color online) GaP optical uncoupled ($T1$) phonon $n=1$ and $n=2$
shell modes at $k_{z}=0$, for a GaAs-GaP core-shell nanowire with fixed $a=3$
nm, as a function of the ratio $\protect\gamma=b/a$. Panel (a) shows the
results without strain effects, and panel (b) including strain. The bulk $TO$
phonon frequency for GaP is indicated by a gray (red) dashed line.}
\label{modosdesacopladosshelln12}\includegraphics[width=%
\columnwidth]{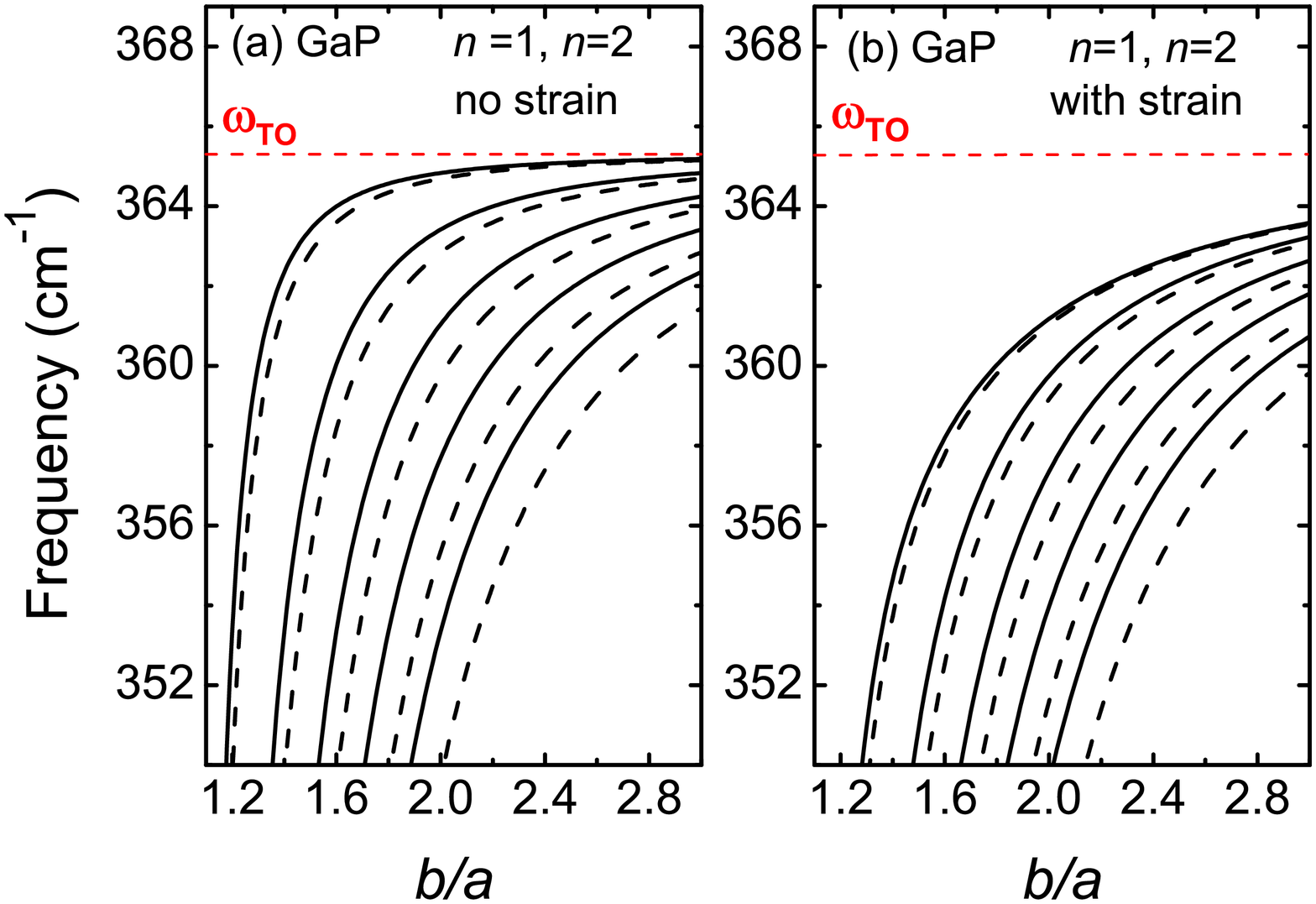}
\end{figure}

The secular equation for the uncoupled $T1$ shell modes takes the form

\begin{eqnarray}  \label{modoT1shellk0}
J_{n}(\mu_{n}^{(m)})N_{n}(\gamma \mu_{n}^{(m)})-J_{n}(\gamma
\mu_{n}^{(m)})N_{n}(\mu_{n}^{(m)})=0,
\end{eqnarray}

with the dispersion relation $\omega ^{2}(\gamma )=\omega _{TO}^{2}-(\mu
_{n}^{(m)}\beta _{T}/a)^{2}$, where the frequencies and parabolicity
parameters correspond to the shell case
An essential difference is observed with respect to core case, namely, that
the mode frequency depends on both core and shell radius.
This dependence is shown in Fig. \ref{modosdesacopladosshelln12} for the
modes $n=1,2$. It is possible to show that the eigenvalues $\mu
_{n}^{(m)}\rightarrow 0$ as $\gamma \rightarrow \infty $, and because the
shift $\Delta \omega _{TO}(\gamma \rightarrow \infty )\rightarrow 0$ (see
Eq.~(\ref{limtrain})), the confined modes tend to the corresponding bulk
$TO$ phonon frequency, i.e., that of GaP.

\begin{figure}[htb]
\includegraphics[width=\columnwidth]{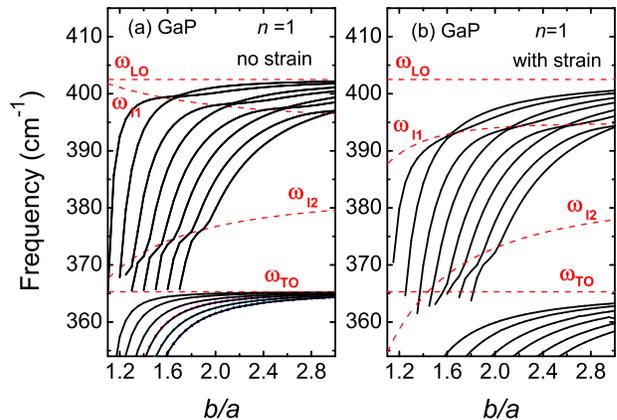}
\caption{(Color online) GaP optical coupled phonon modes at $k_{z}=0$ for $%
n=1$ in GaAs-GaP core-shell nanowire as a function of the relation $\protect%
\gamma=b/a$. Panel (a) neglecting strain, panel (b) considering strain
effects. In the calculation we fixed the value of $a=3$ nm and $\protect%
\varepsilon_{D}=2.56$. The bulk LO and TO phonon frequencies are indicated
by dashed lines. The corresponding I-phonon frequencies obtained in the
framework of the DCA are also represented by a gray (red) dashed line.}
\label{modosacopladosshellradiofijo}
\end{figure}

The boundary conditions (\ref{matchingcond}) for the shell modes yields a
set of equations for the coupled $L$-$T2$ modes, which we give in detail in
the Appendix.
From Eqs. (\ref{modosLT2shellk0}) we obtain the phonon dispersion relation
(unstrained and strained cases) with $\gamma =b/a$ for $n=1$ and $k_{z}=0$,
shown in Fig. \ref{modosacopladosshellradiofijo}.
Notice the two interface shell branches $I1$ and $I2$ (shown by gray(red)
dashed lines), solutions of Eq.~(IF). For frequencies near the interface
phonons $\omega _{I1}$ and $\omega _{I2}$, there is a remarkable mixing
between the longitudinal and transversal modes. This effect is more
remarkable for phonon frequencies $\omega $ near $\omega _{I2}$, where the
anticrossing between two modes with different symmetry is stronger if
compared with the upper interface branch $I1$. Fig.~\ref%
{modosacopladosshellradiofijo} shows that the interface strain pushes down
the phonon shell modes with respect to the bulk phonon frequencies. Recall
that in the core the effect is the opposite (see Fig. \ref%
{modosacopladoscoreespesorfijon1}). Noting the limit (\ref%
{limtrain}), where $\mathrm{tr}(\varepsilon _{s})\rightarrow 0$ for $\gamma
>>1$, it can be seen from Fig. \ref{modosacopladosshellradiofijo} that the
confined modes with $\omega _{L(T)}^{m}$ ($m=1,2,..$) approach the
unstrained $\omega _{LO(TO)}$ bulk phonon frequencies.

\begin{figure}[htb]
\includegraphics[width=\columnwidth]{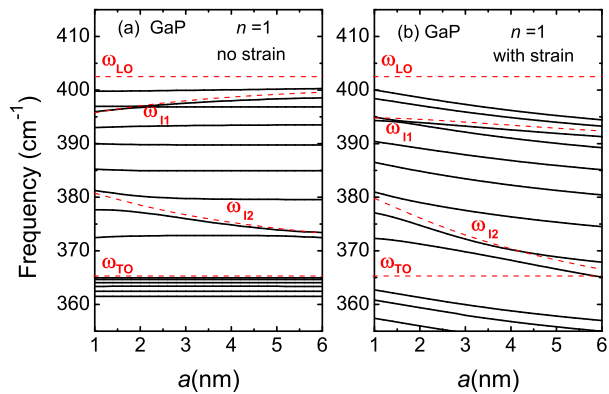}
\caption{(Color online)
The same as
 Fig.~\ref{modosacopladosshellradiofijo}
but as
a
function
of
the
core
radius
$a$,
and fixing
the value
of
$b-a=3$
nm.
}
\label{modosacopladosshellespesorfijo}
\end{figure}

Figure \ref{modosacopladosshellespesorfijo} displays the coupled 
shell modes at $k_{z}=0$ for $n=1$, as a function of the core radius.
Notice that shell modes depend very weakly on the radius $a$ when strain
effects are neglected.
Only four modes close to the I-phonon do have a certain dispersion
(see Fig.~\ref{modosacopladosshellespesorfijo} (a)). When strain effects are
taken into account, there is a general 
dependence in the core radius $a$ for all the modes due to mixing, Fig.~\ref%
{modosacopladosshellespesorfijo}(b). As we fixed $b-a=3$ nm, the shell modes
remain confined even for very large core radius $a$. Moreover, the strain
effect produces 
the shifts $\Delta\omega_{i}=-2\gamma_{i}\omega_{i}\varepsilon_{\mathrm{%
misfit}}(1-2\nu_{c})/(\nu_{c}-1)$ ($i=LO,TO$) to the longitudinal and
transversal confined phonon frequencies.

\subsection{Modes with $k_{z}\neq0$}

Although we can obtain the general expressions for $k_{z}\neq0$, the
equations for the eigenmodes are so lengthy that we only give here those
simpler case, with $n=0$. The form of the basis vectors (\ref{vectoresbase})
for $n=0$, $k_{z}\neq0$ allows us to infer that the modes $T2$ are uncoupled
from the rest.

\subsubsection{Core modes}


The uncoupled transverse $T2$ modes for $n=0$ are given by
$J_{1}(\mu^{(m)}_{1})=0$, which yields the equations for the
eigenmodes 
$\omega_{T}^{2}=\left(\omega_{TO}+\Delta
\omega_{TO}\right)^{2}-\beta_{T}^{2}\left((\mu^{(m)}_{1}/{a}%
)^{2}-k^{2}_{z}\right)$, with $m=1,2,...$. This dispersion relation is equal
to that of the bulk core material, save an energy shift given by the term %
$\left(\mu^{(m)}_{1}\beta_{T}/{a}%
\right)^{2}$, which takes into account the spatial confinement.

\begin{figure}[htb]
\includegraphics[width=\columnwidth]{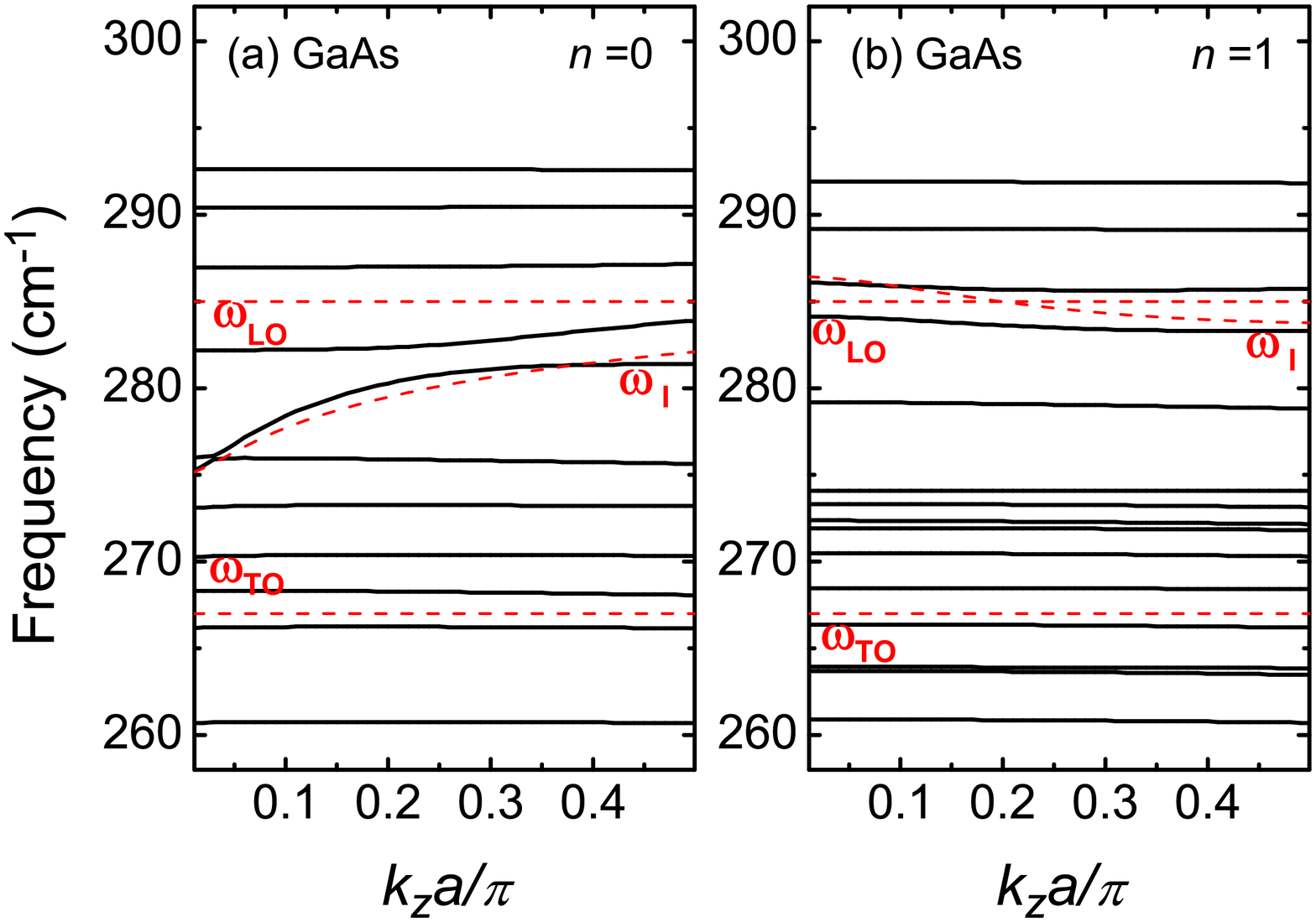} 
\caption{(Color online) Dispersion relations for the core modes of a
core-shell GaAs-GaP nanowire. Panel (a) shows the coupled modes with $n=0$,
panel (b) all modes with $n=1$. $LO$, $TO$ bulk and the $n=1$ interface phonon
frequencies are shown with gray (red) dashed lines. Calculations are for $%
a=3 $ nm and $b/a=2$.}
\label{corekz}
\end{figure}

The system of equations for the $n=0$ coupled $L$-$T1$ core modes can be
written in a rather compact form, as given in the Appendix (Eq.~\ref%
{modosLT1coren0}). From this expression, one can obtain the corresponding
dispersion relations. We do not give the explicit equations for the $n=1$
modes, but we show here the numerical solutions corresponding to these
phonon branches. Fig.~\ref{corekz}(a) depicts the coupled $n=0$ core modes
vs. $k_z$ for a range within a fifth of the first Brillouin zone of the bulk
material, as obtained from Eq.~(\ref{modosLT1coren0}). Fig.~\ref{corekz} (b)
shows the core modes with $n=1$. In both cases we have considered strained
core-shell nanowires.

\subsubsection{Shell modes}

We give here the explicit equations for the simpler case, $n=0$.
The $n=0$ shell transversal $T2$ eigenmodes are obtained from the equation
\begin{eqnarray}  \label{modoT2shelln0}
J_{1}(\mu_{1}^{(m)})N_{1}(\gamma \mu_{1}^{(m)})-J_{1}(\gamma
\mu_{1}^{(m)})N_{1}(\mu_{1}^{(m)})=0,
\end{eqnarray}
which gives $\omega_{T}^{2}=\left(\omega_{TO}+\Delta
\omega_{TO}\right)^{2}-\beta_{T}^{2}\left((\mu^{(m)}_{1}/{a}%
)^{2}-k^{2}_{z}\right)$ with $m=1,2,...$, 
where $\mu_{1}^{(m)}$ represents the solutions of Eq. (\ref{modoT2shelln0})
for a given $\gamma$.


With respect to the coupled $L$-$T1$ shell modes, the
corresponding system of equations, obtained from the boundary conditions for
$n=0$, is given in the Appendix (Eq.~\ref{modosLT1shelln0}). The solution of
these equations yields the dispersion relations presented in Fig. \ref%
{shellkz}(a); in panel (b) we also give the modes for $n=1$.
\begin{figure}[htb]
\includegraphics[width=\columnwidth]{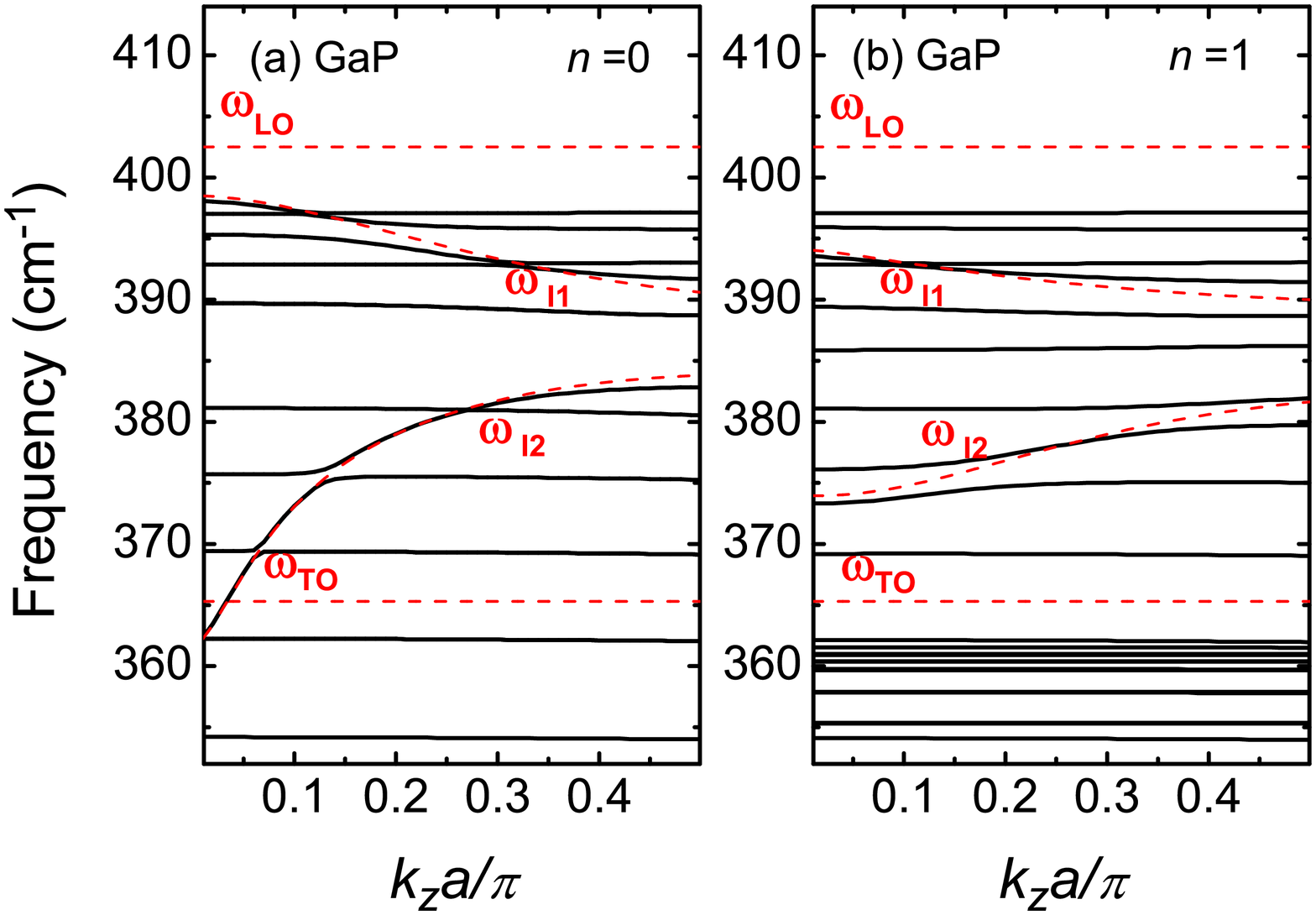} 
\caption{(Color online) Shell modes of a core-shell GaAs-GaP nanowire as a
function of wave vector $k_{z}$. Panel (a) shows the coupled modes with $n=0$%
, panel (b) the $n=1$ modes. $LO$, $TO$ bulk and $n=1$ interface frequencies are
shown with gray (red) dashed lines. Calculations are for fixed $a=3$ nm and $%
b/a=2$.}
\label{shellkz}
\end{figure}

\section{Conclusions}

\label{Conclusions}

In this work we study the confined and interface polar optical phonons in
core-shell nanowires by using a phenomenological continuum approach that
takes into account the coupling of electromechanical oscillations and valid
in the long-wave limit.

We derive a general analytical basis for the oscillations in cylindrical
geometry with circular cross section, which allows for the obtention of
polar vibrational properties in a variety of nanowire systems and materials,
by the appropriate choice of boundary conditions. This permits an
unambiguous identification of the coupled and uncoupled core
and shell modes in terms of the phonon quantum numbers in these novel
structures.

Our results for core and shell modes are summarized as
follows. 
(i) There are uncoupled confined longitudinal and two transversal modes with
$k_{z}=0$, $n=0$; (ii) we have found uncoupled confined transversal modes
with $k_{z}=0$, $n=0,1,2..$ or $k_{z}\neq 0,n=0$; (iii) the modes with $%
k_{z}=0$,$n=0,1,2,...$, have a longitudinal and transversal mixed character,
which couple the mechanical displacement vector $\vec{u}$ and the
electrostatic potential $\varphi $; (iv) there are coupled longitudinal and
two transversal modes for $k_{z}\neq 0,n\neq 0.$ We have also calculated the
interface optical phonons 
in the framework of DCA, involving the electric potential of the phonon
oscillations. We display the dispersion curves for the frequencies and
compared both methods, i.e., the phenomenological continuum approach and the
dielectric model. We have chosen the GaAs-GaP core-shell nanowire as an
example system to apply our model, taking into account strain. We have found
that the inclusion of strain is in fact crucial to model these nanowires,
for their vibrational properties are modified dramatically by such strain
effects. In particular, translationally
invariant $(k_{z}=0$) core modes, which are independent of the shell radius
when strain is neglected, acquire a dependence on the shell thickness if
strain is included. In general, the inclusion of strain produces a shift on
the core and shell modes. While for the core modes the frequencies have an
upward shift, the shell modes present an important downshift of the spectra.
Quantum confinement effects are important for smaller system sizes: when the
core radius $a$ or the shell thickness $b-a$ are smaller than 3 nm, they
become of the order of strain effects. In the core confinement effects have
an opposite sign to those related to strain, whereas in the shell both have
the same tendency.

One of the crucial results of the present work is straightforward
implementation of the Fr\"{o}hlich-like electron-phonon interaction
Hamiltonian, $\widehat{H}_{F}=e\widehat{\varphi },$ for the core-shell
nanowires. Using the general basic expression for the basis
vectors (\ref{basic}) and employing Eqs.~(\ref{genesol}) with the
appropriate boundary conditions (\ref{matchingcond}), we are able to
obtain the eigenfrequencies $\omega _{n,m,k_{z}}$ and the eigensolutions $%
F_{n,m,k_{z}}(\rho )\exp i(n\theta +k_{z}z)$,
which fulfill the
orthonormalization condition
\begin{equation*}
\int\limits_{V}\rho _{M}(\rho )\vec{u}_{n,m,k_{z}}(\rho )\vec{u}_{n^{\prime
},m,k_{z}}(\rho )\rho d\rho =\delta _{nn^{\prime }}.
\end{equation*}%
Thus, we can construct the general solution for the displacement vector $%
\vec{u}(\rho ,\theta ,z)$ and the electrostatic potential $\varphi (\rho
,\theta ,z)$ which in second quantization read

\begin{widetext}
\begin{equation*}
\widehat{\vec{u}}=\sum\limits_{n,m,k_{z}}C_{n,m,k_{z}}\left[ \vec{u}%
_{n,m,k_{z}}(\rho )\exp i(n\theta +k_{z}z)\widehat{b}_{n,m,k_{z}}+H.c.\right]
\end{equation*}%
and
\begin{equation*}
\widehat{\varphi }=\sum\limits_{n,m,k_{z}}C_{n,m,k_{z}}\left[ \varphi
_{n,m,k_{z}}(\rho )\exp i(n\theta +k_{z}z)\widehat{b}_{n,m,k_{z}}+H.c.\right]
,
\end{equation*}
\end{widetext}
where $\widehat{b}_{n,m,k_{z}}$ ($\widehat{b}_{n,m,k_{z}}^{\text{ }+})$ is
the phonon annihilation (creation) operator and the $C_{n,m,k_{z}}$
coefficients are determined by the commutation rules $\left[ \vec{u}(\vec{r}%
),\vec{\pi}(\vec{r}^{\prime })\right] =i\hbar \delta (r-r^{\prime })$ with $%
\vec{\pi}(\vec{r})$ being the momentum conjugate. Thus, it is possible to
show that~\cite{CTG92}
\begin{equation*}
C_{n,m,k_{z}}=\sqrt{\frac{\hbar }{2\omega _{n,m,k_{z}}}}
\end{equation*}%
and the normalized Fr\"{o}hlich interaction Hamiltonian can be cast as
\begin{widetext}
\begin{equation}
\widehat{H}_{F}=\sum\limits_{n,m,k_{z}}e\sqrt{\frac{\hbar }{2\omega
_{n,m,k_{z}}}}\left[ \varphi _{n,m,k_{z}}(\rho )\exp i(n\theta +k_{z}z)%
\widehat{b}_{n,m,k_{z}}+C.C.\right] .  \label{Frolich}
\end{equation}%
\end{widetext}

In the same way, we can argue for the electron-phonon deformation potential
interaction $\widehat{H}_{DP}.$ Notice that the procedure here implemented
allows us to get the electron-phonon interactions that take into account the
spatial confinement effect, the electrostatic influence on the phonon modes
due the interfaces and the strain effect of the core-shell nanowires. Our
work is relevant for the spectroscopic characterization of core-shell
nanowires; it can be of interest for the experimental identification of
these nanostructures.

\acknowledgments

L. C. acknowledges the hospitality of Universidad Aut\'{o}noma del Estado de
Morelos, M\'{e}xico, where this work was envisaged and calculations have
been done. D. G. S.-P. acknowledges CONACyT support.
L. Chico acknowledges financial support of the Spanish MCINN through Grant
FIS2012-33521.

\appendix*

\section{Secular equations for coupled modes}

For the sake of completeness, we give here the lengthier systems of
equations for coupled modes which were omitted in the main text.

The $L$-$T2$ shell modes for $n\neq0$, $k_z=0$ are given by
\begin{widetext}
\begin{eqnarray}
\label{modosLT2shellk0}
\left(
\begin{array}{cccccc}
\frac{n}{t_{s}}J_{n}(t_{s})               & \frac{n}{t_{s}}N_{n}(t_{s})                & J'_{n}(l_{s})               & N'_{n}(l_{s})                & 1 & -1 \\
J'_{n}(t_{s})                             & N'_{n}(t_{s})                              & \frac{n}{l_{s}}J_{n}(l_{s}) & \frac{n}{l_{s}}N_{n}(l_{s})  & 1 & 1 \\
\frac{n}{\gamma t_{s}}J_{n}(\gamma t_{s}) & \frac{in}{\gamma t_{s}}N_{n}(\gamma t_{s}) & J'_{n}(\gamma l_{s})        & N'_{n}(\gamma l_{s})         & \gamma^{n-1} & -\gamma^{-(n+1)} \\
J'_{n}(\gamma t_{s})                      & N'_{n}(\gamma t_{s})                       & \frac{in}{\gamma l_{s}}J_{n}(\gamma l_{s}) & \frac{n}{\gamma l_{s}}N_{n}(\gamma l_{s}) & \gamma^{n-1} & \gamma^{-(n+1)} \\
0                                         & 0 & S_{53}(l_{s}) & S_{54}(l_{s})  & S_{55}(t_{s}) & S_{56}(t_{s}) \\
0                                         & 0 & S_{63}(l_{s}) & S_{64}(l_{s})  & S_{65}(t_{s}) & S_{66}(t_{s}) \\
\end{array}
\right)\left(
              \begin{array}{c}
                A^{(s)}_{T2} \\
                B^{(s)}_{T2} \\
                A^{(s)}_{L}  \\
                B^{(s)}_{L}  \\
                A^{(s)}_{H}  \\
                B^{(s)}_{H}  \\
              \end{array}
            \right)=\left(
              \begin{array}{c}
                0 \\
                0 \\
                0 \\
                0 \\
                0 \\
                0 \\
              \end{array}
            \right)
\end{eqnarray}

\noindent with

 \begin{eqnarray}
 S_{53}(l_{s})&=&J'_{n}(l_{s})-\frac{\varepsilon^{c}_{\infty}n}{\varepsilon^{s}_{\infty}l_{s}}J_{n}(l_{s}),\nonumber S_{54}(l_{s})=N'_{n}(l_{s})-\frac{\varepsilon^{c}_{\infty}n}{\varepsilon^{s}_{\infty}l_{s}}N_{n}(l_{s}), \nonumber\\
 S_{63}(l_{s})&=&J'_{n}(\gamma l_{s})+\frac{\varepsilon_{D}n}{\varepsilon^{s}_{\infty}\gamma l_{s}}J_{n}(\gamma l_{s}),\nonumber
 S_{64}(l_{s})=N'_{n}(\gamma l_{s})+\frac{\varepsilon_{D}n}{\varepsilon^{s}_{\infty}\gamma l_{s}}N_{n}(\gamma l_{s}), \\
 S_{55}(t_{s})&=&\frac{(\varepsilon^{c}_{\infty}-\varepsilon^{s}_{\infty})}{(\varepsilon^{s}_{0}-\varepsilon^{s}_{\infty})}\frac{(\omega^{2}_{TO}-\omega^{2})}{\omega^{2}_{TO}},\nonumber S_{56}(t_{s})=-\frac{(\varepsilon^{c}_{\infty}+\varepsilon^{s}_{\infty})}{(\varepsilon^{s}_{0}-\varepsilon^{s}_{\infty})}\frac{(\omega^{2}_{TO}-\omega^{2})}{\omega^{2}_{TO}},\nonumber\\
 S_{65}(t_{s})&=&-\frac{(\varepsilon_{D}+\varepsilon^{s}_{\infty})}{(\varepsilon^{s}_{0}-\varepsilon^{s}_{\infty})}\frac{(\omega^{2}_{TO}-\omega^{2})}{\omega^{2}_{TO}}\gamma^{n-1},\nonumber
 S_{66}(t_{s})=\frac{(\varepsilon_{D}-\varepsilon^{s}_{\infty})}{(\varepsilon^{s}_{0}-\varepsilon^{s}_{\infty})}\frac{(\omega^{2}_{TO}-\omega^{2})}{\omega^{2}_{TO}}\gamma^{-(n+1)};  \nonumber
 \end{eqnarray}
\end{widetext}

\noindent where $t_{s}=q_{T}a$, $l_{s}=q_{L}a$. In these parameters, as in $%
\omega_{TO}$, it is understood that shell values should be used.

We only give the equations for the dispersion relations, i.e., solutions for
$k_z\neq 0$, for the $n=0$ coupled modes. The core $L$ -$T1$ coupled phonon
bands are given by

\begin{widetext}
 \begin{eqnarray}
\label{modosLT1coren0}
\left(
  \begin{array}{ccccc}
    \frac{k_{a}}{l_{c}}J'_{0}(t_{c}) & J'_{0}(l_{c})                   & I'_{0}(k_{a})                                                      & 0
    & 0 \\                                         -J_{0}(t_{c})                    & \frac{k_{a}}{l_{c}}J_{0}(l_{c}) & I_{0}(k_{a})
    & 0                                     & 0 \\
    0                                & \frac{k_{a}}{l_{c}}J_{0}(l_{c}) & -(1-\frac{\omega^{2}}{\omega^{2}_{TO}})I_{0}(k_{a})  &
    \varepsilon^{c}_{\infty}I_{0}(k_{a})   & \varepsilon^{c}_{\infty}K_{0}(k_{a})  \\
    0                                &  J'_{0}(l_{c})                  & -(1-\frac{\omega^{2}}{\omega^{2}_{TO}})I'_{0}(k_{a}) &
    \varepsilon^{s}_{\infty} I'_{0}(k_{a}) & \varepsilon^{s}_{\infty}K'_{0}(k_{a}) \\
    0                                & 0                               & 0                                                                  & C_{54}
    & C_{55} \\
  \end{array}
\right)\left(
              \begin{array}{c}
                A^{(c)}_{T1}  \\
                A^{(c)}_{L}  \\
                A^{(s)}_{H}  \\
                A^{(s)}_{H}  \\
                B^{(s)}_{H}  \\
              \end{array}
            \right)=\left(
              \begin{array}{c}
                0 \\
                0 \\
                0 \\
                0 \\
                0 \\
              \end{array}
            \right)
\end{eqnarray}
\noindent where

\begin{eqnarray}
C_{54}&=&\varepsilon_{D} I_{0}(\gamma k_{a})K'_{0}(\gamma k_{a})- \varepsilon^{s}_{\infty} I'_{0}(\gamma k_{a})K_{0}(\gamma k_{a}),\\
C_{55}&=& (\varepsilon_{D}-\varepsilon^{s}_{\infty})K_{0}(\gamma k_{a})K'_{0}(\gamma k_{a}).\nonumber
\end{eqnarray}
\end{widetext}

The phonon dispersion relations for the $n=0$ coupled $L$-$T1$ shell modes
can be obtained from

\begin{widetext}
\begin{eqnarray}
\label{modosLT1shelln0}
\left(
  \begin{array}{cccccc}
    \frac{k_{a}}{t_{s}}J'_{0}(t_{s}) & \frac{k_{a}}{t_{s}}N'_{0}(t_{s}) & J'_{0}(l_{s}) & N'_{0}(l_{s}) & I'_{0}(k_{a}) & K'_{0}(k_{a}) \\
    -J_{0}(t_{s}) & -N_{0}(t_{s}) & \frac{k_{a}}{l_{s}}J_{0}(l_{s}) & \frac{k_{a}}{l_{s}}N_{0}(l_{s}) & I_{0}(k_{a}) & K_{0}(k_{a}) \\
    \frac{k_{a}}{t_{s}}J'_{0}(\gamma t_{s}) & \frac{k_{a}}{t_{s}}N'_{0}(\gamma t_{s}) & J'_{0}(\gamma l_{s}) & N'_{0}(\gamma l_{s}) & I'_{0}(\gamma k_{a}) &
    K'_{0}(\gamma k_{a}) \\
    -J_{0}(\gamma t_{s}) & -N_{0}(\gamma t_{s}) & \frac{k_{a}}{l_{s}}J_{0}(\gamma l_{s}) & \frac{k_{a}}{l_{s}}N_{0}(\gamma l_{s}) & I_{0}(\gamma k_{a}) &
    K_{0}(\gamma k_{a}) \\
    0 & 0 & C_{53} & C_{54} & C_{55} & C_{56} \\
    0 & 0 & C_{63} & C_{64} & C_{65} & C_{66} \\
  \end{array}
\right)\left(
              \begin{array}{c}
                A^{(s)}_{T1} \\
                B^{(s)}_{T1} \\
                A^{(s)}_{L}  \\
                B^{(s)}_{L}  \\
                A^{(s)}_{H}  \\
                B^{(s)}_{H}  \\
              \end{array}
            \right)=\left(
              \begin{array}{c}
                0 \\
                0 \\
                0 \\
                0 \\
                0 \\
                0 \\
              \end{array}
            \right)
\end{eqnarray}

\noindent with

\begin{eqnarray}
C_{53}&=&J'_{0}(l_{s})I_{0}(k_{a})-\frac{\varepsilon^{c}_{\infty}}{\varepsilon^{s}_{\infty}}\frac{k_{a}}{l_{s}}J_{0}(l_{s})I'_{0}(k_{a}), \\ \nonumber
C_{54}&=&N'_{0}(l_{s})I_{0}(k_{a})-\frac{\varepsilon^{c}_{\infty}}{\varepsilon^{s}_{\infty}}\frac{k_{a}}{l_{s}}N_{0}(l_{s})I'_{0}(k_{a}), \\ \nonumber
C_{55}&=&\frac{(\varepsilon^{c}_{\infty}-\varepsilon^{s}_{\infty})}{(\varepsilon^{s}_{0}-\varepsilon^{s}_{\infty})}\frac{(\omega^{2}_{TO}-\omega^{2})}{\omega^{2}_{TO}}I_{0}(k_{a})I'_{0}(k_{a}),
\\ \nonumber
C_{56}&=&\frac{(\omega^{2}_{TO}-\omega^{2})}{(\varepsilon^{s}_{0}-\varepsilon^{s}_{\infty})\omega^{2}_{TO}}(\varepsilon^{c}_{\infty}I'_{0}(k_{a})K_{0}(k_{a})-\varepsilon^{s}_{\infty}I_{0}(k_{a})K'_{0}(k_{a})),\nonumber
\\ \nonumber
C_{63}&=&J'_{0}(l_{s})K_{0}(k_{a})-\frac{\varepsilon_{D}}{\varepsilon^{s}_{\infty}}\frac{k_{a}}{l_{s}}J_{0}(l_{s})K'_{0}(k_{a}), \\ \nonumber
C_{64}&=&N'_{0}(l_{s})K_{0}(k_{a})-\frac{\varepsilon_{D}}{\varepsilon^{s}_{\infty}}\frac{k_{a}}{l_{s}}N_{0}(l_{s})K'_{0}(k_{a}), \\ \nonumber
C_{65}&=&\frac{(\omega^{2}_{TO}-\omega^{2})}{(\varepsilon^{s}_{0}-\varepsilon^{s}_{\infty})\omega^{2}_{TO}}(\varepsilon_{D}K'_{0}(\gamma k_{a})I_{0}(\gamma
k_{a})-\varepsilon^{s}_{\infty}K_{0}(\gamma k_{a})I'_{0}(\gamma k_{a})), \\ \nonumber
C_{66}&=&\frac{(\varepsilon_{D}-\varepsilon^{s}_{\infty})}{(\varepsilon^{s}_{0}-\varepsilon^{s}_{\infty})}\frac{(\omega^{2}_{TO}-\omega^{2})}{\omega^{2}_{TO}}K_{0}(\gamma
k_{a})K'_{0}(\gamma k_{a}).\nonumber
\end{eqnarray}
\end{widetext}




\end{document}